\documentclass[sigconf]{acmart}

\usepackage{booktabs} 
\usepackage[english]{babel}
\usepackage{subcaption}
\usepackage{mathrsfs}

\usepackage{multirow}
\usepackage{multicol}
\usepackage{array}

\usepackage{pgfplots}
\usepackage{siunitx}
\usepackage{enumitem}
\usepackage{url}

\setcopyright{rightsretained}

\newcommand{\ms}[1]{}
  \newcommand{\yd}[1]{}
  \newcommand{\cw}[1]{}
  \newcommand{\me}[1]{}
  \newcommand{\hz}[1]{}

\copyrightyear{2017}
\acmYear{2017}

\author{Markus Schedl}
\affiliation{Johannes Kepler University\\Linz, Austria}
\email{markus.schedl@jku.at}

\author{Hamed Zamani}
\affiliation{University of Massachusetts\\Amherst, USA}
\email{zamani@cs.umass.edu}

\author{Ching-Wei Chen}
\affiliation{Spotify USA Inc.\\New York City, USA}
\email{cw@spotify.com}

\author{Yashar Deldjoo}
\affiliation{Politecnico di Milano\\Milan, Italy}
\email{yashar.deldjoo@polimi.it}

\author{Mehdi Elahi}
\affiliation{Free University of Bozen-Bolzano\\Bolzano, Italy}
\email{meelahi@unibz.it}

\begin{document}
\title{Current Challenges and Visions in\\Music Recommender Systems Research}\thanks{This research was supported 
in part by the Center for Intelligent Information Retrieval. Any opinions, findings and conclusions or recommendations expressed in this material are those of the authors and do not necessarily reflect those of the sponsors.}

\begin{abstract}
Music recommender systems (MRS) have ex\-per\-i\-enced a boom in recent years, thanks to the emergence and success of online streaming services, which nowadays make available almost all music in the world at the user's fingertip.
While today's MRS considerably help users to find interesting music in these huge catalogs, MRS research is still facing substantial challenges. In particular when it comes to build, incorporate, and evaluate recommendation strategies that integrate information beyond simple user--item interactions or content-based descriptors, but dig deep into the very essence of listener needs, preferences, and intentions, MRS research becomes a big endeavor and related publications quite sparse.

The purpose of this trends and survey article is twofold. We first identify and shed light on what we believe are the most pressing challenges MRS research is facing, from both academic and industry perspectives. We review the state of the art towards solving these challenges and discuss its limitations.
Second, we detail possible future directions and visions we contemplate for the further evolution of the field. The article should therefore serve two purposes: giving the interested reader an overview of current challenges in MRS research and providing guidance for young researchers by identifying interesting, yet under-researched, directions in the field.
 
\end{abstract}

\maketitle

\section{Introduction}
Research in music recommender systems (MRS) has recently experienced a substantial gain in interest both in academia and industry~\cite{schedl_etal:rsh:2015}. Thanks to music streaming services like Spotify, Pandora, or Apple Music, music aficionados are nowadays given access to tens of millions music pieces. By filtering this abundance of music items, thereby limiting choice overload~\cite{bollen:recsys:2010}, MRS are often very successful to suggest songs that fit their users' preferences. However, such systems are still far from being perfect and frequently produce unsatisfactory recommendations.
This is partly because of the fact that users' tastes and musical needs are highly dependent on a multitude of factors, which are not considered in sufficient depth in current MRS approaches, which are typically centered on the core concept of user--item interactions, or sometimes content-based item descriptors.
In contrast, we argue that satisfying the users' musical entertainment needs requires taking into account intrinsic, extrinsic, and contextual aspects of the listeners~\cite{adomavicius_etal:ai:2011}, as well as more decent interaction information. For instance, personality and emotional state of the listeners (intrinsic) \cite{Rentfrow2003,ferwerda_etal:umap:2015} as well as their activity (extrinsic)~\cite{schedl_etal:mmm:2015,wang_etal:acmmm:2012} are known to influence musical tastes and needs.
So are users' contextual factors including weather conditions, social surrounding, or places of interest~\cite{kaminskas_etal:recsys:2013,adomavicius_etal:ai:2011}.
Also the composition and annotation of a music playlist or a listening session reveals information about which songs go well together or are suited for a certain occasion \cite{mcfee_lanckriet:ismir:2012,zheleva_etal:www:2010}.
Therefore, researchers and designers of MRS should reconsider their users in a holistic way in order to build systems tailored to the specificities of each user.

Against this background, in this trends and survey article, we elaborate on what we believe to be amongst the most pressing current challenges in MRS research, by discussing the respective state of the art and its restrictions (Section~\ref{sec:challenges}).
Not being able to touch all challenges exhaustively, we focus on \textit{cold start}, \textit{automatic playlist continuation}, and \textit{evaluation} of MRS. While these problems are to some extent prevalent in other recommendation domains too, certain characteristics of music pose particular challenges in these contexts. Among them are the short duration of items (compared to movies), the high emotional connotation of music, and the acceptance of users for duplicate recommendations.
In the second part, we present our visions for future directions in MRS research (Section~\ref{sec:visions}). More precisely, we elaborate on the topics of \textit{psychologically-inspired music recommendation} (considering human personality and emotion), \textit{situation-aware music recommendation}, and \textit{culture-aware music recommendation}.
We conclude this article with a summary and identification of possible starting points for the interested researcher to face the discussed challenges (Section~\ref{sec:conclusions}).

The composition of the authors allows to take academic as well as industrial perspectives, which are both reflected in this article.
Furthermore, we would like to highlight that particularly the ideas presented as \emph{Challenge~2: Automatic playlist continuation} in Section~\ref{sec:challenges} play an important role in the task definition, organization, and execution of the ACM Recommender Systems Challenge 2018\footnote{\url{http://www.recsyschallenge.com/2018}} which focuses on this use case. This article may therefore also serve as an entry point for potential participants in this challenge.


\section{Grand Challenges}\label{sec:challenges}
In the following, we identify and detail a selection of the grand challenges, which we believe the research field of music recommender systems is currently facing, i.e., overcoming the cold start problem, automatic playlist continuation, and properly evaluating music recommender systems. We review the state of the art of the respective tasks and its current limitations.

\subsection{Particularities of music recommendation}\label{sec:particular}
Before we start digging deeper into these challenges, we would first like to highlight the major aspects that make music recommendation a particular endeavor and distinguishes it from recommending other items, such as movies, books, or products. These aspects have been adapted and extended from a tutorial on music recommender systems~\cite{schedl_etal:recsys:2017}, co-presented by one of the authors at the ACM Recommender Systems 2017 conference.\footnote{\url{http://www.cp.jku.at/tutorials/mrs_recsys_2017}}

\textit{Duration of items:} In traditional movie recommendation, the items of interest have a typical duration of 90 minutes or more. In book recommendation, the consumption time is commonly even much longer.
In contrast, the duration of music items usually ranges between 3 and 5 minutes (except maybe for classical music). Because of this, music items may be considered more disposable.

\textit{Magnitude of items:} 
The size of common commercial music catalogs is in the range of tens of millions music pieces while movie streaming services have to deal with much smaller catalog sizes, typically thousands up to tens of thousands of movies and series.\footnote{Spotify reports about 30 million songs in 2017 (\url{https://press.spotify.com/at/about}); Amazon's advanced search for books reports 10 million hardcover and 30 million paperback books in 2017 (\url{https://www.amazon.com/Advanced-Search-Books/b?node=241582011}); whereas Netflix, in contrast, offers about 5,500 movies and TV series as of 2016 (\url{http://time.com/4272360/the-number-of-movies-on-netflix-is-dropping-fast}).} 
Scalability is therefore a much more important issue in music recommendation than in movie recommendation. 

\textit{Sequential consumption:}
Unlike movies, music pieces are most frequently consumed sequentially, more than one at a time, i.e., in a listening session or playlist. This yields a number of challenges for a MRS, which relate to identifying the right arrangement of items in a recommendation list.

\textit{Recommendation of previously recommended items:} 
Recommending the same music piece again, at a later point in time, may be appreciated by the user of a MRS, in contrast to a movie or product recommender, where repeated recommendations are usually not preferred.

\textit{Consumption behavior:}
Music is often consumed passively, in the background. While this is not a problem per se, it can affect preference elicitation. In particular when using implicit feedback to infer listener preferences, the fact that a listener is not paying attention to the music (therefore, e.g., not skipping a song) might be wrongly interpreted as a positive signal.

\ms{REWORKED}

\textit{Listening intent and purpose:}
Music serves various purposes for people and hence shapes their intent to listen to it. This should be taken into account when building a MRS. 
In extensive literature and empirical studies, Sch\"{a}fer et al.~\cite{schafer13fp} distilled three fundamental intents of music listening out of 129 distinct music uses and functions: {self-awareness}, {social relatedness}, and {arousal and mood regulation}. 
\emph{Self-awareness} is considered as a very private relationship with music listening. The self-awareness dimension ``helps people think about who they are, who they would like to be, and how to cut their own path''~\cite{Schaefer2017265}. \emph{Social relatedness}~\cite{doi:10.1177/1029864915622054} describes the use of music to feel close to friends and to express identity and values to others. \emph{Mood regulation} is concerned with managing emotions, which is a critical issue when it comes to the well-being of humans~\cite{gross07book,koole09,tami11er}. 
In fact, several studies found that mood and emotion regulation is the most important purpose why people listen to music~\cite{schafer13fp,Lonsdale2011Why,boer10psy,JuslinSloboda2001}, for which reason we discuss the particular role emotions play when listening to music separately below.

\textit{Emotions:} 
Music is known to evoke very strong emotions.\footnote{Please note that the terms ``emotion'' and ``mood'' have different meanings in psychology, whereas they are commonly used as synonyms in music information retrieval (MIR) and recommender systems research. In psychology, in contrast, ``emotion'' refers to a short-time reaction to a particular stimulus, whereas ``mood'' refers to a longer lasting state without relation to a specific stimulus.} 
This is a mutual relationship, though, since also the emotions of users affect musical preferences~\cite{bodner07,punkanen11jad,gross07book}.
Due to this strong relationship between music and emotions, the problem of automatically describing music in terms of emotion words is an active research area, commonly refereed to as music emotion recognition (MER), e.g.~\cite{YangChen2012,barthet12cmmr,kim10ismir}.
Even though MER can be used to tag music by emotion terms, how to integrate this information into MRS is a highly complicated task, for three reasons. First, MER approaches commonly neglect the distinction between \emph{intended emotion} (i.e., the emotion the composer, songwriter, or performer had in mind when creating or performing the piece), \emph{perceived emotion} (i.e., the emotion recognized while listening), and \emph{induced emotion} that is felt by the listener.
Second, the preference for a certain kind of emotionally laden music piece depends on whether the user wants to enhance or to modulate her mood.
Third, emotional changes often occur within the same music piece, whereas tags are commonly extracted for the whole piece.
Matching music and listeners in terms of emotions therefore requires to model the listener's musical preference as a time-dependent function of their emotional experiences, also considering the intended purpose (mood enhancement or regulation).
This is a highly challenging task and usually neglected in current MRS, for which reason we discuss emotion-aware MRS as one of the main future directions in MRS research, cf.~Section~\ref{sec:psychology}.

\textit{Listening context:}
Situational or contextual aspects \cite{RN16,RN52} have a strong influence on music preference, consumption, and interaction behavior.
For instance, a listener will likely create a different playlist when preparing for a romantic dinner than when warming-up with friends to go out on a Friday night~\cite{schedl_etal:mmm:2015}. 
The most frequently considered types of context include location (e.g., listening at workplace, when commuting, or relaxing at home)~\cite{kaminskas_etal:recsys:2013} and time (typically categorized into, e.g., morning, afternoon, and evening)~\cite{cebrian_etal:recsys:2010}. 
Context may, in addition, also relate to the listener's activity~\cite{wang_etal:acmmm:2012}, weather~\cite{pettijohnmusic}, or the use of different listening devices, e.g., earplugs on a smartphone vs.~hi-fi stereo at home~\cite{schedl_etal:mmm:2015}, to name a few.
Since music listening is also a highly social activity, investigating the social context of the listeners is crucial to understand their listening preferences and behavior~\cite{cunningham_nichols:ismir:2009,ohara:springer:2006}.
The importance of considering such contextual factors in MRS research is acknowledged by discussing situation-aware MRS as a trending research direction, cf.~Section~\ref{sec:situation}.

\ms{REWORKED END}


\subsection{Challenge 1: Cold start problem}

\paragraph*{\textbf{Problem definition:}} $\;$
One of the major problems of recommender systems in general \cite{elahi2016survey,rubens2015active}, and music recommender systems in particular  \cite{li2005probabilistic,kaminskas2012contextual} is the {\it cold start} problem, i.e., when a new user registers to the system or a new item is added to the catalog 
and the system does not have sufficient data associated with these items/users. In such a case, the system cannot properly recommend existing items to a new user 
({\it new user} problem) or recommend a new item to the existing users 
({\it new item} problem) \cite{elahi14ec-web-survey,kaminskas2012contextual,adomaviciustowards,shein02}. 

Another sub-problem of cold start is the {\it sparsity} problem which refers to the fact that the number of given ratings is much lower than the number of possible ratings, which is particularly likely when the number of users and items is large. The inverse of the ratio between given and possible ratings is called sparsity. High sparsity translates into low rating \emph{coverage}, since most users tend to rate only a tiny fraction of items. The effect is that recommendations often become  unreliable~\cite{kaminskas2012contextual}.
Typical values of sparsity are quite close to 100\% in most real-world recommender systems. In the music domain, this is a particularly substantial problem. Dror et al.~\cite{dror2011yahoo}, for instance, analyzed the Yahoo! Music dataset, which as of time of writing represents the largest music recommendation dataset. They report a sparsity of $99.96\%$. For comparison, the Netflix dataset of movies has a sparsity of ``only'' $98.82\%$.\footnote{Note that Dror et al.'s analysis was conducted in 2011. Even though the general character (rating matrices for music items being sparser than those of movie items) remained the same, the actual numbers for today's catalogs are likely slightly different.}

\paragraph*{\textbf{State of the art:}} $\;$


A number of approaches have already been proposed to tackle the cold start problem in the music recommendation domain, foremost content-based approaches, hybridization, cross-domain recommendation, and active learning. 

\emph{Content-based recommendation} (CB) algorithms do not require ratings of users other than the target user. Therefore, as long as some pieces of information about the user's own preferences are available, such techniques can be used in cold start scenarios. Furthermore, in the most severe case, when a new item is added to the catalog, content-based methods enable recommendations, because they can extract features from the new item and use them to make recommendations. It is noteworthy that while collaborative filtering (CF) systems have cold start problems both for new users and new items, content-based systems have only cold start problems for new users~\cite{aggarwal2016content}.
\\
As for the new item problem, a standard approach is to extract a number of features that define the acoustic properties of the audio signal and use content-based learning of the user interest (user profile learning) in order to effect recommendations. 
Feature extraction is typically done automatically, but can also be effected manually by musical experts, as in the case of Pandora's Music Genome Project.\footnote{\url{http://www.pandora.com/about/mgp}}
Pandora uses up to 450 specific descriptors per song, such as ``aggressive female vocalist'', ``prominent backup vocals'', ``abstract lyrics'', or ``use of unusual harmonies''.\footnote{\url{http://enacademic.com/dic.nsf/enwiki/3224302}}
Regardless of whether the feature extraction process is performed automatically or manually, this approach is advantageous not only to address the new item problem but also because an accurate feature representation can be highly predicative of users' tastes and interests which can be leveraged in the subsequent information filtering stage~\cite{aggarwal2016content}. An advantage of music to video is that features in music is limited to a single audio channel, compared to audio and visual channels for videos adding a level complexity to the content analysis of videos explored individually or multimodal in different research works~\cite{mei2011contextual,deldjoo2016content,elahi2017exploring,deldjoo2017effect}.


Automatic feature extraction from audio signals can be done in two main manners: (1) by extracting a feature vector from each item individually, independent of other items or (2) by considering the cross-relation between items in the training dataset. The difference is that in (1) the same process is performed in the training and testing phases of the system, and the extracted feature vectors can be used off-the-shelf in the subsequent processing stage, for example they can be used to compute similarities between items in a one-to-one fashion at testing time. In contrast, in (2) first a model is built from all features extracted in the training phase, whose main role is to map the features into a new (acoustic) space in which the similarities between items are better represented and exploited. 
An example of approach (1) is the block-level feature framework~\cite{seyerlehner_etal:smc:2010,seyerlehner_etal:ismir:2010}, which creates a feature vector of about 10,000 dimensions, independently for each song in the given music collection. This vector describes aspects such as spectral patterns, recurring beats, and correlations between frequency bands.
An example of strategy (2) is to create a low-dimensional i-vector representation from the Mel-frequency cepstral coefficients (MFCCs), which model musical timbre to some extent~\cite{eghbal-zadeh_etal:ismir:2015}.
To this end, a universal background model is created from the MFCC vectors of the whole music collection, using a Gaussian mixture model (GMM). Performing factor analysis on a representation of the GMM eventually yields i-vectors.

In scenarios where some form of semantic labels, e.g., genres or musical instruments, are available, 
it is possible to build models that learn the intermediate mapping between low-level audio features and semantic representations using machine learning techniques, and subsequently use the learned models for prediction. A good point of reference for such \emph{semantic-inferred} approaches can be found in~\cite{cheng2016effective,bogdanov_etal:ipm:2013}.

An alternative technique to tackle the new item problem is \emph{hybridization}. A review of different hybrid and ensemble recommender systems can be found in~\cite{burke2002hybrid,aggarwal2016ensemble}. In~\cite{donaldson:recsys:2007} the authors propose a music recommender system which combines an acoustic CB and an item-based CF recommmender. 
For the content-based component, it computes acoustic features including spectral properties, timbre, rhythm, and pitch. 
The content-based component then assists the collaborative filtering recommender in tackling the cold start problem since the features of the former are automatically derived via audio content analysis.
\\
The solution proposed in~\cite{yoshii2006hybrid} is a hybrid recommender system that combines CF and acoustic CB strategies also by feature hybridization. However, in this work the feature-level hybridization is not performed in the original feature domain. Instead, a set of latent variables referred to as \textit{conceptual genre} are introduced, whose role is to provide a common shared feature space for the two recommenders and enable hybridization. The weights associated with the latent variables reflect the musical taste of the target user and are learned during the training stage.
\\
In~\cite{shao_etal:taslp:2009} the authors propose a hybrid recommender system incorporating item--item CF and acoustic CB based on similarity metric learning. The proposed metric learning is an optimization model that aims to learn the weights associated with the audio content features (when combined in a linear fashion) so that a degree of consistency between CF-based similarity and the acoustic CB similarity measure is established. The optimization problem can be solved using quadratic programming techniques.

Another solution to cold start are \emph{cross-domain recommendation} techniques, which aim at improving recommendations in one domain (here music) by making use of information about the user preferences in an auxiliary domain~\cite{fernandez2016alleviating,cantador14tutorial}. Hence, the knowledge of the preferences of the user is transferred from an auxiliary domain to the music domain, resulting in a more complete and accurate user model.
Similarly, it is also possible to integrate additional pieces of information about the (new) users, which are not directly related to music, such as their personality, in order to improve the estimation of the user's music preferences. Several studies conducted on user personality characteristics support the conjecture that it may be useful to exploit this information 
in music recommender systems~\cite{north2008social,Rentfrow2003,HuP10study,ferwerda_etal:empire:2016,ferwerda_etal:chi:2015}. 
For a more detailed literature review of cross-domain recommendation, we refer to~\cite{fernandez2012cross,Cantador2015,khan2017cross}.

In addition to the aforementioned approaches, {\it active learning} has shown promising results in dealing with the cold start problem in single domain ~\cite{Rashid08learningpref,ElahiRR11} or cross-domain recommendation scenario \cite{zhang2016multi,PaganoQEC17}. Active learning addresses this problem at its origin by identifying and eliciting (high quality) data that can represent the preferences of users better than by what they provide themselves. Such a system therefore interactively demands specific user feedback to maximize the improvement of system performance.

  
\paragraph*{\textbf{Limitations:}} $\;$
The state-of-the-art approaches elaborated on above are restricted by certain limitations.
When using \textit{content-based filtering}, for instance, almost all existing approaches rely on a number of predefined audio features that have been used over and over again, including spectral features, MFCCs, and a great number of derivatives~\cite{knees2016music}. However, doing so assumes that (all) these features are predictive of the user's music taste, while in practice it has been shown that the acoustic properties that are important for the perception of music are highly subjective~\cite{novello_etal:ismir:2006}. Furthermore, listeners' different tastes and levels of interest in different pieces of music influence perception of item similarity~\cite{schedl_etal:jiis:2013}.
This subjectiveness demands for CB recommenders that incorporate personalization in their mathematical model. For example, in~\cite{elbadrawy2015user} the authors propose a hybrid (CB+CF) recommender model, namely regression-based latent factor models (RLFM). 
In \cite{agarwal2009regression} the authors propose a user-specific feature-based similarity model (UFSM), which defines a similarity function for each user, leading to a high degree of personalization. Although not designed specifically for the music domain, the authors of \cite{agarwal2009regression} provide an interesting literature review of similar user-specific models. 


While \textit{hybridization} can therefore alleviate the cold start problem to a certain extent, as seen in the examples above, respective approaches are often complex, computationally expensive, and lack transparency~\cite{Burke2007}. In particular, results of hybrids employing latent factor models are typically hard to understand for humans.

A major problem with \textit{cross-domain recommender systems} is their need for data that connects two or more target domains, e.g., books, movies, and music~\cite{Cantador2015}.
In order for such approaches to work properly, items, users, or both therefore need to overlap to a certain degree~\cite{cremonesi:icdmw:2011}.
In the absence of such overlap, relationships between the domains must be established otherwise, e.g.,~by inferring semantic relationships between items in different domains or assuming similar rating patterns of users in the involved domains. However, whether respective approaches are capable of transferring knowledge between domains is disputed~\cite{Cremonesi:2014:CRW:2645710.2645769}. 
A related issue in cross-domain recommendation is that there is a lack of established datasets with clear definitions of domains and recommendation scenarios~\cite{khan2017cross}. Because of this, the majority of existing work on cross-domain RS use some type of conventional recommendation dataset transformation to suit it for their need.

Finally, also \textit{active learning} techniques suffer from a number of issues.
First of all, the typical active learning techniques propose to a user to rate the items that the system has predicted to be interesting for them, i.e., the items with highest predicted ratings. This indeed is a default strategy in recommender systems for eliciting ratings since users tend to rate what has been recommended to them. Even when users browse the item catalog, they are more likely to rate items which they like or are interested in, rather than those items that they dislike or are indifferent to. Indeed, it has been shown that 
doing so creates a strong bias in the collected rating data as the database gets populated   disproportionately with high ratings. This in turn may substantially influence the prediction algorithm and decrease the recommendation accuracy~\cite{elahi14active}. 

Moreover, not all the active learning strategies are necessarily personalized. The users differ very much in the amount of information they have about the items, their preferences, and the way they make decisions. Hence, it is clearly inefficient to request all the users to rate the same set of items, because many users may 
have a very limited knowledge, ignore many items, and will therefore not provide ratings for these items. 
Properly designed active learning techniques should take this into account and  propose different items to different users to rate. This can be highly  beneficial and increase the chance of acquiring ratings of higher quality~\cite{elahi2011adaptive}.  

Moreover, the traditional interaction model designed for active learning in recommender systems can support building the initial profile of a user mainly in the sign-up process. This is done by generating a user profile by requesting the user to rate a set of selected items~\cite{Carenini2003Toward}. On the other hand, the users must be able to also update their profile by providing more ratings anytime they are willing to. This requires the system to adopt a {conversational} interaction model~\cite{Carenini2003Toward}, e.g., by exploiting novel interactive design elements in the user interface~\cite{Cremonesi2017}, such as explanations that can describe the benefits of providing more ratings and motivating the user to do so.  

Finally, it is important to note that  in an up-and-running recommender system, the ratings are given by users not only when requested by the system (active learning) but also when a user voluntarily explores the item catalog and rates some familiar items (natural acquisition of ratings) \cite{McNee03IEN,Carenini2003Toward,Rashid08learningpref,elahi12adapting,elahi14active}. While this could have a huge impact on the performance of the system, it has been mostly ignored by the  majority of the research works in the field of active learning for recommender systems. Indeed, almost all research works have been based on a rather non-realistic assumption that the only source for collecting new ratings is through the system requests. Therefore, it is crucial to take into account a more realistic scenario when studying the active learning techniques in recommender systems, which can better picture how the system evolves over time when ratings are provided by users \cite{Rashid08learningpref,pu2012evaluating}.

\subsection{Challenge 2: Automatic playlist continuation}

\paragraph*{\textbf{Problem definition:}} $\;$
In its most generic definition, a playlist is simply a sequence of tracks intended to be listened to together. The task of automatic playlist generation (APG) then refers to the automated creation of these sequences of tracks.
\ms{CHANGED: } In this context, the ordering of songs in a playlist to generate is often highlighted as a characteristics of APG, which is a highly complex endeavor. Some authors have therefore proposed approaches based on Markov chains to model the transitions between songs in playlists, e.g.~\cite{chen_etal:kdd:2012,mcfee_lanckriet:ismir:2011}. While these approaches have been shown to outperform approaches agnostic of the song order in terms of log likelihood, recent research has found little evidence that the exact order of songs actually matters to users~\cite{Tintarev:2017:SDS:3079628.3079633}, while the ensemble of songs in a playlist~\cite{vall_etal:recsys:2017} and direct song-to-song transitions~\cite{kamehkhosh_etal:milc:2018} do matter.
\ms{CHANGED END}

Considered a variation of APG, the task of \textit{automatic playlist continuation} (APC) consists of adding one or more tracks to a playlist in a way that fits the same target characteristics of the original playlist. This has benefits in both the listening and creation of playlists: users can enjoy listening to continuous sessions beyond the end of a finite-length playlist, while also finding it easier to create longer, more compelling playlists without needing to have extensive musical familiarity.

A large part of the APC task is to accurately infer the intended purpose of a given playlist. This is challenging not only because of the broad range of these intended purposes (when they even exist), but also because of the diversity in the underlying features or characteristics that might be needed to infer those purposes. 

Related to Challenge 1, an extreme cold start scenario for this task is where a playlist is created with some metadata (e.g., the title of a playlist), but no song has been added to the playlist. This problem can be cast as an \emph{ad-hoc information retrieval task}, where the task is to rank songs in response to a user-provided metadata query. 

The APC task can also potentially benefit from user profiling, e.g., making use of previous playlists and the long-term listening history of the user. We call this \emph{personalized playlist continuation}. 

According to a study carried out in 2016 by the Music Business Association\footnote{\url{https://musicbiz.org/news/playlists-overtake-albums-listenership-says-loop-study}} as part of their Music Biz Consumer Insights program,\footnote{\url{https://musicbiz.org/resources/tools/music-biz-consumer-insights/consumer-insights-portal}} 
playlists accounted for 31\% of music listening time among listeners in the USA, more than albums (22\%), but less than single tracks (46\%). 
Other studies, conducted by MIDiA,\footnote{\url{https://www.midiaresearch.com/blog/announcing-midias-state-of-the-streaming-nation-2-report}} 
show that 55\% of streaming music service subscribers create music playlists, with some streaming services such as Spotify currently hosting over 2 billion playlists.\footnote{\url{https://press.spotify.com/us/about}} 
In a 2017 study conducted by Nielsen,\footnote{\url{http://www.nielsen.com/us/en/insights/reports/2017/music-360-2017-highlights.html}} it was found that 58\% of users in the USA create their own playlists, 32\% share them with others. 
Studies like these suggest a growing importance of playlists as a mode of music consumption, and as such, the study of APG and APC has never been more relevant.

\paragraph*{\textbf{State of the art:}} $\;$
APG has been studied ever since digital multimedia transmission made huge catalogs of music available to users. Bonnin and Jannach provide a comprehensive survey of this field in~\cite{bonnin2015}. In it, the authors frame the APG task as the creation of a sequence of tracks that fulfill some ``target characteristics'' of a playlist, given some ``background knowledge'' of the characteristics of the catalog of tracks from which the playlist tracks are drawn. Existing APG systems tackle both of these problems in many different ways.

In early approaches~\cite{pachet_etal:icmcs:1999,alghoniemy_etal:acmmm:2000,alghoniemy_tewfik:icme:2001} the \emph{target characteristics} of the playlist are specified as multiple explicit constraints, which include musical attributes or metadata such as artist, tempo, and style. In others, the target characteristics are a single seed track~\cite{logan:ismir:2002} or a start and an end track~\cite{chen_etal:kdd:2012,flexer_etal:ismir:2008,alghoniemy_tewfik:icme:2001}.
Other approaches create a circular playlist that comprises all tracks in a given music collection, in such a way that consecutive songs are as similar as possible~\cite{pohle_etal:trmm:2007,knees_etal:mir:2006}.
In other works, playlists are created based on the context of the listener, either as single source~\cite{schedl_etal:icmr:2014} or in combination with content-based similarity~\cite{cheng_shen:icmr:2014,reynolds_etal:am:2007}.

A common approach to build the \emph{background knowledge} of the music catalog for playlist generation is using machine learning techniques to extract that knowledge from manually-curated playlists. The assumption here is that curators of these playlists are encoding rich latent information about which tracks go together to create a satisfying listening experience for an intended purpose. Some proposed APG and APC systems are trained on playlists from sources such as online radio stations~\cite{maillet:ismir:2009,chen_etal:kdd:2012}, online playlist websites~\cite{mcfee_lanckriet:ismir:2012,vall_etal:recsys:2017}, and music streaming services~\cite{pichl_etal:icdmw:2015}. In the study by Pichl et al.~\cite{pichl_etal:icdmw:2015}, the names of playlists on Spotify were analyzed to create contextual clusters, which were then used to improve recommendations. 

An approach to specifically address song ordering within playlists is the use of generative models that are trained on hand-curated playlists. McFee and Lanckriet~\cite{mcfee_lanckriet:ismir:2011} represent songs by metadata, familiarity, and audio content features, adopting ideas from statistical natural language processing. They train various Markov chains to model transitions between songs.
Similarly, Chen et al.~\cite{chen_etal:kdd:2012} propose a logistic Markov embedding to model song transitions. This is similar to matrix decomposition methods and results in an embedding of songs in Euclidean space. In contrast to McFee and Lanckriet's model, Chen et al.'s model does not use any audio features.


\paragraph*{\textbf{Limitations:}} $\;$
While some work on automated playlist continuation highlights the special characteristics of playlists, i.e.,~their \textit{sequential order}, it is not well understood to which extent and in which cases taking into account the order of tracks in playlists helps create better models for recommendation. For instance, in~\cite{vall_etal:recsys:2017}~Vall et al.~recently demonstrated on two datasets of hand-curated playlists that the song order seems to be negligible for accurate playlist continuation when a lot of popular songs are present. On the other hand, the authors argue that order does matter when creating playlists with tracks from the long tail. Another study by McFee and Lanckriet~\cite{mcfee_lanckriet:ismir:2012} also suggests that transition effects play an important role in modeling playlist continuity.
This is in line with a study presented by Kamehkhosh et al.~in~\cite{kamehkhosh_etal:milc:2018}, in which users identified song order as being the second but last important criterion for playlist quality.\footnote{The ranking of criteria (from most to least important) was: homogeneity, artist diversity, transition, popularity, lyrics, order, and freshness.}
In another recent user study~\cite{Tintarev:2017:SDS:3079628.3079633} conducted by Tintarev et al., the authors found that many participants did not care about the order of tracks in recommended playlists, sometimes they did not even notice that there is a particular order. However, this study was restricted to 20 participants who used the Discover Weekly service of Spotify.\footnote{\url{https://www.spotify.com/discoverweekly}}

Another challenge for APC is evaluation: in other words, how to assess the quality of a playlist. Evaluation in general is discussed in more detail in the next section, but there are specific questions around evaluation of playlists that should be pointed out here. As Bonnin and Jannach~\cite{bonnin2015} put it, the ultimate criterion for this is \textit{user satisfaction}, but that is not easy to measure. In~\cite{mcfee_lanckriet:ismir:2011}, McFee and Lanckriet categorize the main approaches to APG evaluation as human evaluation, semantic cohesion, and sequence prediction. Human evaluation comes closest to measuring user satisfaction directly, but suffers from problems of scale and reproducibility. Semantic cohesion as a quality metric is easily measurable and reproducible, but assumes that users prefer playlists where tracks are similar along a particular semantic dimension, which may not always be true, see for instance the studies carried out by Slaney and White~\cite{slaney2006} and by Lee~\cite{lee:ismir:2011}. 
Sequence prediction casts APC as an information retrieval task, but in the domain of music, an inaccurate prediction need not be a bad recommendation, and this again leads to a potential disconnect between this metric and the ultimate criterion of user satisfaction.

Investigating which factors are potentially important for a positive user perception of a playlist, Lee conducted a qualitative user study~\cite{lee:ismir:2011}, investigating playlists that had been automatically created based on content-based \textit{similarity}. They made several interesting observations.
A concern frequently raised by participants was that of consecutive songs being too similar, and a general lack of \textit{variety}. However, different people had different interpretations of variety, e.g., variety in genres or styles vs.~different artists in the playlist. Similarly, different criteria were mentioned when listeners judged the coherence of songs in a playlist, including lyrical content, tempo, and mood.
When creating playlists, participants mentioned that similar lyrics, a common theme (e.g., music to listen to in the train), story (e.g., music for the Independence Day), or era (e.g., rock music from the 1980s) are important and that tracks not complying negatively effect the flow of the playlist.
These aspects can be extended by responses of participants in a study conducted by Cunningham et al.~\cite{cunningham_etal:ismir:2006}, who further identified the following categories of playlists: same artist, genre, style, or orchestration, playlists for a certain event or activity (e.g., party or holiday), romance (e.g., love songs or breakup songs), playlists intended to send a message to their recipient (e.g., protest songs), and challenges or puzzles (e.g., cover songs liked more than the original or songs whose title contains a question mark). 

Lee also found that \textit{personal preferences} play a major role. In fact, already a single song that is very much liked or hated by a listener can have a strong influence on how they judge the entire playlist~\cite{lee:ismir:2011}. This seems particularly true if it is a highly disliked song~\cite{cunningham_etal:ismir:2005}.
Furthermore, a good \textit{mix of familiar and unknown songs} was often mentioned as an important requirement for a good playlist. 
Supporting the discovery of interesting new songs, still contextualized by familiar ones, increases the likelihood of realizing a \textit{serendipitous encounter} in a playlist~\cite{zhang_etal:wsdm:2012,schedl_etal:carr:2012}.
Finally, participants also reported that their familiarity with a playlist's genre or theme influenced their judgment of its quality. In general, listeners were more picky about playlists whose tracks they were familiar with or they liked a lot. 


Supported by the studies summarized above, we argue that the question of what makes a great playlist is highly subjective and further depends on the intent of the creator or listener. Important criteria when creating or judging a playlist include track similarity/coherence, variety/diversity, but also the user's personal preferences and familiarity with the tracks, as well as the intention of the playlist creator.
Unfortunately, current automatic approaches to playlist continuation are agnostic of the underlying psychological and sociological factors that influence the decision of which songs users choose to include in a playlist.
Since knowing about such factors is vital to  understand the intent of the playlist creator, we believe that algorithmic methods for APC need to holistically learn such aspects from manually created playlists and integrate respective intent models.
However, we are aware that in today's era where billions of playlists are shared by users of online streaming services,\footnote{\url{https://press.spotify.com/us/about}} a large-scale analysis of psychological and sociological background factors is impossible.
Nevertheless, in the absence of explicit information about user intent, a possible starting point to create intent models might be the metadata associated with user-generated playlists, such as title or description.
To foster this kind of research, the playlists provided in the dataset for the ACM Recommender Systems Challenge 2018 include playlist titles.\footnote{\url{https://recsys-challenge.spotify.com
%https://recsys.acm.org/recsys18/challenge
}} 








\subsection{Challenge 3: Evaluating music recommender systems}

\paragraph*{\textbf{Problem definition:}} $\;$
Having its roots in machine learning (cf.~rating prediction) and information retrieval (cf.~``retrieving'' items based on implicit ``queries'' given by user preferences), the field of recommender systems originally adopted evaluation metrics from these neighboring fields.
In fact, accuracy and related quantitative measures, such as precision, recall, or error measures (between predicted and true ratings), are still the most commonly employed criteria to judge the recommendation quality of a recommender system~\cite{baeza-yates:modern_ir:2011,eval:rsh:2015}.
In addition, novel measures that are tailored to the recommendation problem have emerged in recent years.
These so-called beyond-accuracy measures
~\cite{Kaminskas:2016:DSN:3028254.2926720} address the particularities of recommender systems and gauge, for instance, the utility, novelty, or serendipity of an item. 
However, a major problem with these kinds of measures is that they integrate factors that are hard to describe mathematically, for instance, the aspect of surprise in case of serendipity measures. 
For this reason, there sometimes exist a variety of different definitions to quantify the same beyond-accuracy aspect.

\begin{table*}[t!]
\centering
\begin{tabular}{|l|c|c|c|}
\hline
Measure & Abbreviation & Type & Ranking-aware \\
\hline
Mean absolute error & MAE & error/accuracy & No \\
Root mean square error & RMSE & error/accuracy & No \\
Precision at top K recommendations & P@K & accuracy & No \\
Recall at top K recommendations & R@K & accuracy & No \\
Mean average precision at top K recommendations & MAP@K & accuracy & Yes \\
Normalized discounted cumulative gain & NDCG & accuracy & Yes \\
Half-life utility & HLU & accuracy & Yes \\
Mean percentile rank & MPR & accuracy & Yes \\
Spread & --- & beyond & No \\
Coverage & --- & beyond & No \\
Novelty & --- & beyond & No \\
Serendipity & --- & beyond & No \\
Diversity & --- & beyond & No \\
\hline
\end{tabular}
\caption{\label{tab:eval}Evaluation measures commonly used for recommender systems.}
\end{table*}

\paragraph*{\textbf{State of the art:}} $\;$
In the following, we discuss performance measures which are most frequently reported when evaluating recommender systems. An overview of these is given in Table~\ref{tab:eval}. They can be roughly categorized into accuracy-related measures, such as prediction error (e.g., MAE and RMSE) or standard IR measures (e.g., precision and recall), and beyond-accuracy measures, such as diversity, novelty, and serendipity.
Furthermore, while some of the metrics quantify the ability of recommender systems to find good items, e.g., precision and recall, 
others consider the ranking of items and therefore assess the system's ability to position good recommendations at the top of the recommendation list, e.g., MAP, NDCG, or MPR.

\textit{Mean absolute error (MAE)}
\label{subsec:mae-measure}
 is one of the most common metrics for evaluating the prediction power of recommender algorithms.
It computes the average absolute deviation between the predicted ratings and the
actual ratings provided by users~\cite{herockereval}. Indeed, MAE indicates how close the rating predictions generated by an MRS are to the real user ratings.
MAE is computed as follows:

\begin{equation}
MAE=\frac{1}{|T|}\sum_{r_{u,i} \in T} {|r_{u,i}-\hat{r}_{u,i}|}
\end{equation}
where $r_{u,i}$ and $\hat{r}_{u,i}$ respectively denote the actual and the predicted ratings of item $i$ for user $u$. 
MAE sums over the absolute prediction
errors for all ratings in a test set $T$. 

\textit{Root mean square error (RMSE)} is another similar metric that is computed as: 
\begin{equation}
RMSE=\sqrt{\frac{1}{|T|}\sum_{r_{u,i} \in T} {(r_{u,i}-\hat{r}_{u,i})^2}}
\end{equation}
It is an extension to MAE in that the error term is squared, which penalizes larger differences between predicted and true ratings more than smaller ones. This is motivated by the assumption that, for instance, a rating prediction of 1 when the true rating is 4 is much more severe than a prediction of 3 for the same item.
 
\textit{Precision at top $K$ recommendations (P@K)}
is a common metric that measures the accuracy of the system in commanding relevant items. 
In order to compute $P@K$, for each user, the top $K$ recommended items whose ratings also appear in the test set $T$ are considered. This metric was originally designed for binary relevance judgments. Therefore, in case of availability of relevance information at different levels, such as a five point Likert scale, the labels should be binarized, e.g., considering the ratings greater than or equal to~$4$ (out of~	$5$) as relevant.
For each user $u$, $P_u@K$ is computed as follows: 
\begin{equation}
P_u@K=\frac{|L_u \cap \hat{L}_u| }{|\hat{L}_u|}
\end{equation}
where $L_u$ is the set of relevant items for user $u$ in the test set $T$ and $\hat{L}_u$ denotes the recommended set containing the $K$ items in $T$ with the highest predicted ratings for the user $u$. The overall $P@K$ is then computed by averaging $P_u@K$ values for all users in the test set.

\textit{Mean average precision at top K recommendations (MAP@K)} is a rank-based metric that computes the overall precision of the system at different lengths of recommendation lists.
MAP is computed as the arithmetic mean of the average precision over the entire set of users in the test set. Average precision for the top $K$ recommendations ($AP@K$) is defined as follows:
\begin{equation}
AP@K = \frac{1}{N} \sum_{i=1}^{K} {P@i \, \cdot \, rel(i)}
\end{equation}

where $rel(i)$ is an indicator signaling if the $i^{\mathrm{th}}$ recommended item is relevant, i.e.~$rel(i)=1$, or not, i.e.~$rel(i)=0$; $N$ is the total number of relevant items. 
Note that MAP implicitly incorporates recall, because it also considers the relevant items not in the recommendation list.\footnote{We should note that in the recommender systems community, another variation of average precision is gaining popularity recently, formally defined by: $AP@K = \frac{1}{\min(K,N)} \sum_{i=1}^{K} {P@i \, \cdot \, rel(k)}$ in which $N$ is the total number of relevant items and $K$ is the size of recommendation list. The motivation behind the minimization term is to prevent the AP scores to be unfairly suppressed when the number of recommendations is too low to capture all the relevant items. This variation of MAP was popularized by Kaggle competitions~\cite{kaggle18} about recommender systems and has been used in several other research works, consider for example~\cite{mcfee2012million,aiolli2013efficient}.} 

\textit{Recall at top $K$ recommendations (R@K)} is presented here for the sake of completeness, even though it is not a crucial measure from a consumer's perspective. Indeed, the listener is typically not interested in being recommended all or a large number of relevant items, rather in having good recommendations at the top of the recommendation list.
For a user $u$, $R_u@K$ is defined as:
\begin{equation}
R_u@K = \frac{|L_u \cap \hat{L}_u| }{|L_u|}
\end{equation}
where $L_u$ is the set of relevant items of user $u$ in the test set $T$ and $\hat{L}_u$ denotes the recommended set containing the $K$ items in $T$ with the highest predicted ratings for the user $u$. The overall $R@K$ is calculated by averaging $R_u@K$ values for all the users in the test set.

\textit{Normalized discounted cumulative gain (NDCG)} is a measure for the ranking quality of the recommendations. This metric has originally been proposed to evaluate effectiveness of information retrieval systems~\cite{cgbasedtech}. It is nowadays also frequently used for evaluating music recommender systems~\cite{EigenRankNdcg,Park2009Pairwise,adaptivecf}.
Assuming that the recommendations for user $u$ are
sorted according to the predicted rating values in descending order. $DCG_u$ is defined as follows:
\begin{equation}
DCG_u = \sum_{i=1}^N \frac{r_{u,i}}{log_{2} (i+1)}
\end{equation}
where $r_{u,i}$ is the true rating (as found in test set $T$) for the
item ranked at position $i$ for user $u$, and $N$ is the length of
the recommendation list.
Since the rating distribution depends on the users' behavior, the DCG values for different users are not directly comparable. Therefore, the cumulative gain for each user should be normalized. This is done by computing the ideal DCG for user $u$, denoted as $IDCG_u$, which is the $DCG_u$ value for the best possible ranking, obtained by ordering the items by true ratings in descending order.
Normalized discounted cumulative gain for user $u$ is then
calculated as:
\begin{equation}
NDCG_u = \frac{DCG_u}{IDCG_u}
\end{equation}
Finally, the overall normalized discounted cumulative gain $N\!DCG$ is computed by averaging $N\!DCG_u$ over the
entire set of users.

\vspace{5mm}
In the following, we present common quantitative evaluation metrics, which have been particularly designed or adopted to assess recommender systems performance, even though some of them have their origin in information retrieval and machine learning. The first two (HLU and MRR) still belong to the category of accuracy-related measures, while the subsequent ones capture beyond-accuracy aspects

\textit{Half-life utility (HLU)} measures the utility of a recommendation list for a user with the assumption that the likelihood of viewing/choosing a recommended item by the user exponentially decays with the item's position in the ranking~\cite{breese1998empirical,pan2008one}. 
Formally written, HLU for user $u$ is defined as:
\begin{equation}
HLU_u = \sum_{i=1}^{N}{\frac{\max{(r_{u,i}-d, 0)}}{2^{(rank_{u,i}-1)/(h-1)}}}
\end{equation}
where $r_{u,i}$ and $rank_{u,i}$ denote the rating and the rank of item $i$ for user $u$, respectively, in the recommendation list of length $N$; $d$ represents a default rating (e.g., average rating) and $h$ is the half-time, calculated as the rank of a music item in the list, such that the user can eventually listen to it with a $50\%$ chance. $HLU_u$ can be further normalized by the maximum utility (similar to NDCG), and the final HLU is the average over the half-time utilities obtained for all users in the test set. A larger {HLU} may correspond to a superior recommendation performance. 

\textit{Mean percentile rank (MPR)} estimates the users' satisfaction with items in the recommendation list, and is computed as the average of the percentile rank for each test item within the ranked list of recommended items for each user~\cite{hu2008collaborative}. 
The percentile rank of an item is the percentage of items whose position in the recommendation list is equal to or lower than the position of the item itself.
Formally, the percentile rank $PR_u$ for user $u$ is defined as:
\begin{equation}
PR_u = \frac{ \displaystyle \sum_{r_{u,i} \in T} r_{u,i} \cdot rank_{u,i}}{ \displaystyle \sum_{r_{u,i} \in T} r_{u,i}}
\end{equation}
where $r_{u,i}$ is the true rating (as found in test set $T$) for item $i$ rated by user $u$ and $rank_{u,i}$ is the percentile rank of item $i$ within the ordered list of recommendations for user $u$. MPR is then the arithmetic mean of the individual $PR_u$ values over all users.
A randomly ordered recommendation list has an expected MPR value of $50\%$. A smaller MPR value is therefore assumed to correspond to a superior recommendation performance. 


\textit{Spread} is a metric of how well the recommender algorithm can spread its attention across a larger set of items~\cite{Kluver:2014:ERB:2645710.2645742}. 
In more detail, spread is the entropy of the distribution of the items recommended to the users in the test set. It is formally defined as:
\begin{equation}
spread = -\sum_{i \in I}{P(i) \log{P(i)}}
\end{equation}
where $I$ represents the entirety of items in the dataset and $P(i) = count(i) / \sum_{i' \in I}{count(i')}$, such that $count(i)$ denotes the total number of times that a given item $i$ showed up in the recommendation lists. It may be infeasible to expect an algorithm to achieve the perfect spread (i.e., recommending each item an equal number of times) without avoiding irrelevant recommendations or unfulfillable rating requests. Accordingly, moderate spread values are usually preferable.

\textit{Coverage} of a recommender system is defined as the proportion of items over which the system is capable of generating recommendations~\cite{herockereval}:
\begin{equation}
coverage =  \frac{|\hat T|}{|T|}
\end{equation}
where $|T|$ is the size of the test set and $|\hat T|$ is the number of ratings in $T$ for which the system can predict a value.
This is particularly important in cold start situations, when recommender systems are not able to accurately predict the ratings of new users or new items, and hence obtain low coverage. Recommender systems with lower coverage are therefore limited in the number of items they can recommend.  
A simple remedy to improve low coverage is to implement some default recommendation strategy for an unknown user--item entry. For example, we can consider the average rating of users for an item as an estimate of its rating. This may come at the price of accuracy and therefore the trade-off between coverage and accuracy needs to be considered in the evaluation process~\cite{aggarwal2016evaluating}. 

\textit{Novelty} measures the ability of a recommender system to recommend new items that the user did not know about before~\cite{adamopoulos2015unexpectedness}. A recommendation list may be accurate, but if it contains a lot of items that are not novel to a user, 
it is not necessarily a useful list~\cite{zhang_etal:wsdm:2012}.
\\
While novelty should be defined on an individual user level, considering the actual freshness of the recommended items, it is common to use 
the self-information of the recommended items relative to their global popularity:
\begin{equation}
novelty =\frac{1}{|U|}{\sum_{u \in U}}{\sum_{i \in L_{u}}}\frac{- \log_2{pop_i}}{N}
\end{equation}
where $pop_i$ is the popularity of item $i$ measured as percentage of users who rated $i$, $L_u$ is the recommendation list of the top $N$ recommendations for user $u$~\cite{zhang_etal:wsdm:2012,zhou2010solving}. The above definition assumes that the likelihood of the user selecting a previously unknown item is proportional to its global popularity and is used as an approximation of novelty. 
In order to obtain more accurate information about novelty or freshness, explicit user feedback is needed, in particular since the user might have listened to an item through other channels before.
\\
It is often assumed that the users prefer recommendation lists with more novel items.
However, if the presented items are too novel, then the user is unlikely to have any knowledge of them, nor to be able to understand or rate them. Therefore, moderate values indicate better performances~\cite{Kluver:2014:ERB:2645710.2645742}. 


\textit{Serendipity} aims at evaluating MRS based on the \emph{relevant and surprising} recommendations.
While the need for serendipity is commonly agreed upon~\cite{herlocker_etal:tis:2004}, the question of how to measure the degree of serendipity for a recommendation list is controversial. This particularly holds for the question of whether the factor of surprise implies that items must be novel to the user~\cite{Kaminskas:2016:DSN:3028254.2926720}.
On a general level, serendipity of a recommendation list $L_u$ provided to a user $u$ can be defined as:
\begin{equation}
serendipity(L_u) = \frac{\left| L_u^{unexp} \cap L_u^{useful} \right|} {\left| L_u \right|}
\end{equation}
where 
$L_u^{unexp}$ and $L_u^{useful}$ denote subsets of $L$ that contain, respectively, recommendations unexpected to and useful for the user.
The usefulness of an item is commonly assessed by explicitly asking users or taking user ratings as proxy~\cite{Kaminskas:2016:DSN:3028254.2926720}.
The unexpectedness of an item is typically quantified by some measure of distance from expected items, i.e.,~items that are similar to the items already rated by the user.
In the context of MRS, Zhang et al. \cite{zhang_etal:wsdm:2012} propose an ``unserendipity'' measure that is defined as the average similarity between the items in the user's listening history and the new recommendations. Similarity between two items in this case is calculated by an adapted cosine measure that integrates co-liking information, i.e., number of users who like both items.
It is assumed that lower values correspond to more surprising recommendations, since lower values indicate that recommendations deviate from the user's traditional behavior \cite{zhang_etal:wsdm:2012}.

\textit{Diversity} is another beyond-accuracy measure as already discussed in the limitations part of Challenge 1.
It gauges the extent to which recommended items are different from each other, where difference can relate to various aspects, e.g., musical style, artist, lyrics, or instrumentation, just to name a few.
Similar to serendipity, diversity can be defined in several ways. One of the most common is to compute pairwise distance between all items in the recommendation set, either averaged~\cite{ziegler2005improving} or summed~\cite{Smyth:2001:SVD:646268.758890}.
In the former case, the diversity of a recommendation list $L$ is calculated as follows:
\begin{equation}
diversity(L) = \frac{ \displaystyle \sum_{i \in L} \sum_{j \in L \setminus i} dist_{i,j}}{|L| \cdot \left(|L|-1\right)}
\end{equation}
where $dist_{i,j}$ is the some distance function defined between items $i$ and $j$. Common choices are inverse cosine similarity~\cite{Ribeiro:2012:PHM:2365952.2365962}, inverse Pearson correlation~\cite{vargas_castells:recsys:2011}, or Hamming distance~\cite{Kelly2006}.

\vspace{5mm}
When it comes to the task of evaluating playlist recommendation, where the goal is to assess the capability of the recommender in providing proper transitions between subsequent songs, the conventional error or accuracy metrics may not be able to capture this property. 
There is hence a need for \emph{sequence-aware evaluation} measures.
For example, consider the scenario where a user who likes both classical and rock music is recommended a rock music right after she has listened to a classic piece. Even though both music styles are in agreement with her taste, the transition between songs plays an important role toward user satisfaction. In such a situation, given a currently played song and in presence of several equally likely good options to be played next, a RS may be inclined to rank songs based on their popularity. Hence, other metrics such as \emph{average log-likelihood} have been proposed to better model the transitions~\cite{chen2013multi,chen2012playlist}. In this regard, when the goal is to suggest a sequence of items, alternative \textit{multi-metric} evaluation approaches are required to take into consideration multiple quality factors. Such evaluation metrics can consider the ranking order of the recommendations or the internal coherence or diversity of the recommended list as a whole. In many scenarios, adoption of such quality metrics can lead to a trade-off with accuracy which should be balanced by the RS algorithm~\cite{quadrana2018sequence}.

\paragraph*{\textbf{Limitations:}} $\;$
As of today, the vast majority of evaluation approaches in recommender systems research focuses on quantitative measures, either accuracy-like or beyond-accuracy, which are often computed in offline studies.\\
Doing so has the advantage of facilitating the reproducibility of evaluation results. However, limiting the evaluation to quantitative measures means to forgo another important factor, which is user experience. In other words, in the absence of user-centric evaluations, it is difficult to extend the claims to the more important objective of the recommender system under evaluation, i.e., giving users a pleasant and useful personalized experience~\cite{knijnenburg2015evaluating}.

Despite acknowledging the need for more user-centric evaluation strategies~\cite{schedl_etal:jiis:2013}, the factor human, user, or, in the case of MRS, listener is still way too often neglected or not properly addressed. 
For instance, while there exist quantitative objective measures for serendipity and diversity, as discussed above, \emph{perceived} serendipity and diversity can be highly different from the measured ones~\cite{Vargas:2014:CRS:2645710.2645743} as they are subjective user-specific concepts. 
This illustrates that even beyond-accuracy measures cannot fully capture the real \emph{user satisfaction} with a recommender system. 
On the other hand, approaches that address user experience (UX) can be investigated to evaluate recommender systems. For example, a MRS can be evaluated based on \emph{user engagement}, which provides a restricted explanation of UX that concentrates on judgment of product quality during interaction~\cite{oro46332,Lehmann:2012,O'Brien:2010}.
User satisfaction, user engagement, and more generally user experience are commonly assessed through user studies~\cite{barrington2009smarter,lee2016users,lee2017understanding}.

\begin{figure*}[th]
  \centering
  \includegraphics[width=.8\textwidth]{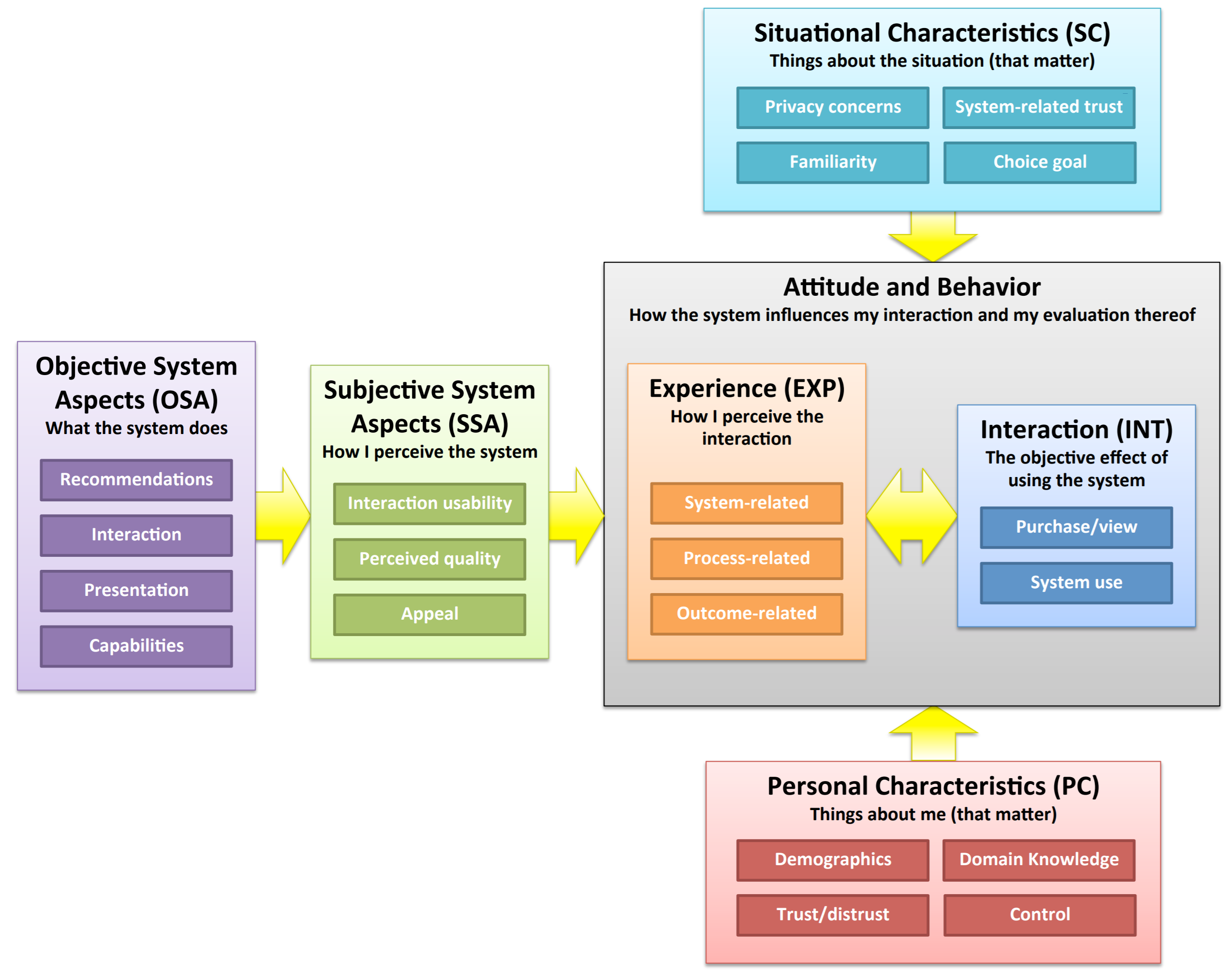}
  \caption{\label{fig:user-centric}Evaluation framework of the user experience for recommender systems, according to~\cite{knijnenburg2012explaining}.}
\end{figure*}

Addressing both objective and subjective evaluation criteria, Knijnenburg et al.~\cite{knijnenburg2012explaining} propose a holistic framework for user-centric evaluation of recommender systems. Figure~\ref{fig:user-centric} provides an overview of the components. The objective system aspects (OSA) are considered unbiased factors of the RS, including aspects of the user interface, computing time of the algorithm, or number of items shown to the user. They are typically easy to specify or compute.
The OSA influence the subjective system aspects (SSA), which are caused by momentary, primary evaluative feelings while interacting with the system~\cite{Hassenzahl2005}. This results in a different perception of the system by different users. SSA are therefore highly individual aspects and typically assessed by user questionnaires. Examples of SSA include general appeal of the system, usability, and perceived recommendation diversity or novelty.
The aspect of experience (EXP) describes the user's attitude towards the system and is commonly also investigated by questionnaires. It addresses the user's perception of the interaction with the system. The experience is highly influenced by the other components, which means changing any of the other components likely results in a change of EXP aspects. 
Experience can be broken down into the evaluation of the system, the decision process, and the final decisions made, i.e., the outcome.	
\ms{CHANGED: } The interaction (INT) aspects describe the observable behavior of the user, time spent viewing an item, as well as clicking or purchasing behavior. In a music context, examples further include liking a song or adding it to a playlist.  
\ms{CHANGED END}
Therefore, interactions aspects belong to the objective measures and are usually determined via logging by the system.
Finally, Knijnenburg et al.'s framework mentions personal characteristics (PC) and situational characteristics (SC), which influence the user experience. PC include aspects that do not exist without the user, such as user demographics, knowledge, or perceived control, while SC include aspects of the interaction context, such as when and where the system is used, or situation-specific trust or privacy concerns.
Knijnenburg et al.~\cite{knijnenburg2012explaining} also propose a questionnaire to asses the factors defined in their framework, for instance, perceived recommendation quality, perceived system effectiveness, perceived recommendation variety, choice satisfaction, intention to provide feedback, general trust in technology, and system-specific privacy concern.

While this framework is a generic one, tailoring it to MRS would allow for user-centric evaluation thereof. Especially the aspects of personal and situational characteristics should be adapted to the particularities of music listeners and listening situations, respectively, cf. Section~\ref{sec:particular}. To this end, researchers in MRS should consider the aspects relevant for the perception and preference of music, and their implications on MRS, which have been identified in several studies, e.g.~\cite{laplante:ismir:2014,schedl_etal:taffc:2017,schedl_etal:jiis:2013,CunninghamEtAl_2007_FindNewMusiA,LaplanteEtAl_2006_EverLifeMusiInfo}.
In addition to the general ones mentioned by Knijnenburg et al., of great importance in the music domain seem to be psychological factors, including affect and personality, social influence, musical training and experience, and physiological condition.

We believe that carefully and holistically evaluating MRS by means of accuracy and beyond-accuracy, objective and subjective measures, in offline and online experiments, would lead to a better understanding of the listeners' needs and requirements vis-\`a-vis MRS, and eventually 
a considerable improvement of current MRS.

\section{Future Directions and Visions}\label{sec:visions}
While the challenges identified in the previous section are already researched on intensely, in the following, we provide a more forward-looking analysis and discuss some MRS-related trending topics, which we assume influential for the next generation of MRS.
All of them have in common that their aim is to create more personalized recommendations.
More precisely, we first outline how psychological constructs such as personality and emotion could be integrated into MRS. Subsequently, we address situation-aware MRS and argue for the need of multifaceted user models that describe contextual and situational preferences.
To round off, we discuss the influence of users' cultural background on recommendation preferences, which needs to be considered when building culture-aware MRS.

\subsection{Psychologically-inspired music recommendation}\label{sec:psychology}
{Personality} and {emotion} are important psychological constructs.
While personality characteristics of humans are a predictable and stable measure that shapes human behaviors, emotions are short-term affective responses to a particular stimulus~\cite{Tkalcic2016}.
Both have been shown to influence music tastes~\cite{ferwerda_etal:umap:2015,Schaefer2017265,schedl_etal:taffc:2017} and user requirements for MRS~\cite{ferwerda_etal:chi:2015,ferwerda_etal:empire:2016}.
However, in the context of (music) recommender systems, personality and emotion do not play a major role yet. Given the strong evidence that both influence listening preferences~\cite{Rentfrow2003,schedl_etal:taffc:2017} and the recent emergence of approaches to accurately predict them from user-generated data~\cite{kosinski_etal:pnas:2013,skowron_etal:www:2016}, we believe that psychologically-inspired MRS is an upcoming area.

\subsubsection{Personality:} In psychology research, personality is often defined as a ``consistent behavior pattern and interpersonal processes originating within the individual''~\cite{burger10personality}. 
This definition accounts for the individual differences in people's emotional, interpersonal, experiential, attitudinal,
and motivational styles~\cite{john99bigfive}.
Several prior works have studied the relation of decision making and personality factors. In \cite{Rentfrow2003}, as an example, it has been shown that personality can influence the 
human decision making process as well as the tastes and interests. 
Due to this direct relation, people with similar personality factors are very likely to share similar interests and tastes.

Earlier studies conducted on the user personality characteristics support the potential benefits that personality information could have in recommender systems  \cite{hu2011Enhancing,hu09comparative,elahi2013personality,marko13ldos,BraunhoferUserPers2015,braunhofer2014techniques,uitdenbogerd2002review}. 
As a known example, psychological studies \cite{Rentfrow2003} have shown that
extravert people are likely to prefer the upbeat and conventional music. 
Accordingly, a personality-based MRS could use this information to better predict which songs are more likely than others to please extravert people~\cite{HuP10study}. Another example of potential usage is to exploit personality information in order to compute similarity among users and hence identify the like-minded users~\cite{marko13ldos}. This similarity information could then be integrated into a neighborhood-based collaborative filtering approach. 

In order to use personality information in a recommender system, the system first has to elicit this information from the users, which can be done either explicitly or implicitly. In the former case, the system can ask the user to complete a personality questionnaire using one of the personality evaluation inventories, e.g., the ten item personality inventory~\cite{gosling2003very} or the big five inventory~\cite{John1999Big}. In the latter case, the system can learn the personality by tracking and observing users' behavioral patterns, for instance, Liking behavior on Facebook~\cite{kosinski_etal:pnas:2013} or applying filters to images posted on Instagram~\cite{skowron_etal:www:2016}.
Not too surprisingly, it has shown that systems that explicitly elicit personality characteristics achieve superior recommendation outcomes, e.g., in terms of user satisfaction, ease of use, and prediction accuracy~\cite{dunn2009Eval}. 
On the downside, however, many users are not willing to fill in long questionnaires before being able to use the RS. A way to alleviate this problem is to ask users only the most informative questions of a personality instrument~\cite{schedl_etal:mmsys:2016}. Which questions are most informative, though, first needs to be determined based on existing user data and is dependent on the recommendation domain.
Other studies showed that users are to some extent willing to provide further information in return for a better quality of recommendations~\cite{swearingen2001beyond}. 

Personality information can be used in various ways, particularly, to generate recommendations when traditional rating or consumption data is missing. Otherwise, the personality traits can be seen as an additional feature that extends the user profile, that can be used mainly to identify similar users in neighborhood-based recommender systems or directly fed into extended matrix factorization models~\cite{fernandez2016alleviating}.

\ms{UPDATED:}
\subsubsection{Emotion:} The emotional state of the MRS user has a strong impact on his or her short-time musical preferences~\cite{kaminskas2012contextual}. Vice versa, music has a strong influence on our emotional state. It therefore does not come as a surprise that emotion regulation was idenfied as one of the main reasons why people listen to music~\cite{Lonsdale2011Why,schafer13fp}.
As an example, people may listen to completely different musical genres or styles when they are sad in comparison to when they are happy. Indeed, prior research on music psychology discovered that people may choose the type of music which moderates their emotional condition~\cite{konecni1982social}. More recent findings show that music can be mainly chosen so as to augment the emotional situation perceived by the listener~\cite{north1996situational}. 
In order to build emotion-aware MRS, it is therefore necessary to (i) infer the emotional state the listener is in, (ii) infer emotional concepts from the music itself, and (iii) understand how these two interrelate. These three tasks are detailed below.

\paragraph{Eliciting the emotional state of the listener:}
Similar to personality traits, the emotional state of a user can be elicited explicitly or implicitly. In the former case, the user is typically presented one of the various \textit{categorical models} (emotions are described by distinct emotion words such as happiness, sadness, anger, or fear)~\cite{zentner2008emotions,hevner:pr:1935} or \textit{dimensional models} (emotions are described by scores with respect to two or three dimensions, e.g., valence and arousal)~\cite{russell:jpsp:1980}. For a more detailed elaboration on emotion models in the context of music, we refer to~\cite{yang_chen:mer:2011,schedl_etal:taffc:2017}.
The implicit acquisition of emotional states can be effected, for instance, by analyzing user-generated text~\cite{6850785}, speech~\cite{Erdal2016}, or facial expressions in video~\cite{EbrahimiKahou:2015:RNN:2818346.2830596}.

\paragraph{Emotion tagging in music:}
The music piece itself can be regarded as an emotion-laden content and in turn can be described by emotion words. 
The task of automatically assigning such emotion words to a music piece is an active research area, often refereed to as music emotion recognition (MER), e.g.~\cite{YangChen2012,barthet12cmmr,kim10ismir,yang_chen:tist:2013,huq_etal:jnmr:2010,zentner2008emotions}.
How to integrate such emotion terms created by MER tools into a MRS is, however, not an easy task, for several reasons.
First, early MER approaches usually neglected the distinction between \emph{intended emotion}, \emph{perceived emotion}, and \emph{induced or felt emotion}, cf.~Section~\ref{sec:particular}. Current MER approaches focus on perceived or induced emotions.
However, musical content still contains various characteristics that affect the emotional state of the listener, such as lyrics, rhythm, and harmony, and the way how they affect the emotional state is highly subjective. This so even though research has detected a few general rules, for instance, a musical piece that is in major key is typically perceived brighter and happier than those in minor key, or a piece in rapid tempo is perceived more exciting or more tense than slow tempo ones~\cite{kuo2005emotion}.

\paragraph{Connecting listener emotions and music emotion tags:}
Current emotion-based MRS typically consider emotional scores as contextual factors that characterize the situation the user is experiencing. Hence, the recommender systems exploit emotions in order to pre-filter the preferences of users or post-filter the generated recommendations.
Unfortunately, this neglects the psychological background, in particular on the subjective and complex interrelationships between expressed, perceived, and induced emotions~\cite{schedl_etal:taffc:2017}, which is of special importance in the music domain as music is known to evoke stronger emotions than, for instance, products~\cite{schedl_etal:recsys:2017}.
It has also been shown that personality influences in which emotional state which kind of emotionally laden music is preferred by listeners~\cite{ferwerda_etal:umap:2015}.
Therefore, even if automated MER approaches would be able to accurately predict the perceived or induced emotion of a given music piece, in the absence of deep psychological listener profiles, matching emotion annotations of items and listeners may not yield satisfying recommendations. This is so because how people judge music and which kind of music they prefer depends to a large extent on their current psychological and cognitive states.
We hence believe that the field of MRS should embrace psychological theories, elicit the respective user-specific traits, and integrate them into recommender systems, in order to build decent emotion-aware MRS.
\ms{UPDATED END.}

 

\subsection{Situation-aware music recommendation}\label{sec:situation}
\ms{UPDATED: }
Most of the existing music recommender systems make recommendations solely based on a set of user-specific and item-specific signals. However, in real-world scenarios, many other signals are available. These additional signals can be further used to improve the recommendation performance. A large subset of these additional signals includes \emph{situational signals}. In more detail, the music preference of a user depends on the situation at the moment of recommendation.\footnote{Please note that music taste is a relatively stable characteristic, while music preferences vary depending on the context and listening intent.}
\emph{Location} is an example of situational signals; for instance, the music preference of a user would differ in libraries and in gyms~\cite{cheng_shen:icmr:2014}. Therefore, considering location as a situation-specific signal could lead to substantial improvements in the recommendation performance. \emph{Time of the day} is another situational signal that could be used for recommendation; for instance, the music a user would like to listen to in mornings differs from those in nights~\cite{cunningham_etal:am:2008}. 
One situational signal of particular importance in the music domain is \emph{social context} since music tastes and consumption behaviors are deeply rooted in the users' social identities and mutually affect each other~\cite{cunningham_nichols:ismir:2009,ohara:springer:2006}. For instance, it is very likely that a user would prefer different music when being alone than when meeting friends. Such social factors should therefore be considered when building situation-aware MRS.
Other situational signals that are sometimes exploited include the user's current \emph{activity}~\cite{wang_etal:acmmm:2012}, the \emph{weather}~\cite{pettijohnmusic}, the user's \emph{mood}~\cite{north2011situational}, and the \emph{day of the week}~\cite{herrera_rocking_2010}. 
Regarding time, there is also another factor to consider, which is that most music that was considered trendy years ago is now considered old. This implies that ratings for the same song or artist might strongly differ, not only between users, but in general as a function of time. To incorporate such aspects in MRS, it would be crucial to record a timestamp for all ratings.

It is worth noting that situational features have been proven to be strong signals in improving retrieval performance in search engines \cite{Bennett:2011,Zamani:2017}. Therefore, we believe that researching and building situation-aware music recommender systems should be one central topic in MRS research. 

While several situation-aware MRS already exist, e.g.~\cite{hi_ogihara:ismir:2011,cheng_shen:icmr:2014,wang_etal:acmmm:2012,schedl_etal:icmr:2014,baltrunas_etal:ecweb:2011,kaminskas_etal:recsys:2013}, they commonly exploit only one or very few such situational signals, or are restricted to a certain usage context, e.g., music consumption in a car or in a tourist scenario.
Those systems that try to take a more comprehensive view and consider a variety of different signals, on the other hand, suffer from a low number of data instances or users, rendering it very hard to build accurate context models~\cite{schedl_etal:mmm:2015}.
What is still missing, in our opinion, are (commercial) systems that integrate a variety of situational signals on a very large scale in order to truly understand the listeners needs and intents in any given situation and recommend music accordingly.
While we are aware that data availability and privacy concerns counteract the realization of such systems on a large commercial scale, we believe that MRS will eventually integrate decent multifaceted user models inferred from contextual and situational factors. 

\subsection{Culture-aware music recommendation}
While most humans share an inclination to listen to music, independent on their location or cultural background, the way music is performed, perceived, and interpreted evolves in a culture-specific manner.
However, research in MRS seems to be agnostic of this fact. 
In music information retrieval (MIR) research, on the other hand, cultural aspects have been studied to some extent in recent years, after preceding (and still ongoing) criticisms of the predominance of Western music in this community.
Arguably the most comprehensive culture-specific research in this domain has been conducted as part of the CompMusic project,\footnote{http://compmusic.upf.edu} in which five non-Western music traditions have been analyzed in detail in order to advance automatic description of music by emphasizing cultural specificity.
The analyzed music traditions included Indian Hindustani and Carnatic~\cite{DBLP:conf/ismir/DuttaM14}, Turkish Makam~\cite{georgi:ismir:2016}, Arab-Andalusian~\cite{sordo_etal:2014}, and Beijing Opera~\cite{repetto_serra:ismir:2014}. 
However, the project's focus was on music creation, content analysis, and ethnomusicological aspects rather than on the music consumption side~\cite{serra:jnmr:2014,serra:aes:2014,doi:10.1080/09298215.2013.812123}.
Recently, analyzing content-based audio features describing rhythm, timbre, harmony, and melody for a corpus of a larger variety of world and folk music with given country information, Panteli et al.~found distinct acoustic patterns of the music created in individual countries~\cite{Panteli_etal:ismir:2016}. 
They also identified geographical and cultural proximities that are reflected in music features, looking at outliers and misclassifications in a classification experiments using country as target class. For instance, Vietnamese music was often confused with Chinese and Japanese, South African with Botswanese.

In contrast to this --- meanwhile quite extensive --- work on culture-specific analysis of music traditions, little effort has been made to analyze cultural differences and patterns of music consumption behavior, which is, as we believe, a crucial step to build culture-aware MRS.
The few studies investigating such cultural differences include~\cite{hu_lee:ismir:2012}, in which Hu and Lee found differences in perception of moods between American and Chinese listeners. By analyzing the music listening behavior of users from 49 countries, Ferwerda et al.~found relationships between music listening diversity and Hofstede's cultural dimensions~\cite{ferwerda_etal:umap:2016,ferwerda_schedl:soap:2016}. Skowron et al.~used the same dimensions to predict genre preferences of listeners with different cultural backgrounds~\cite{skowron2017predicting}.
Schedl analyzed a large corpus of listening histories created by Last.fm users in 47 countries and identified distinct preference patterns~\cite{ijmir:schedl:2017}. Further analyses revealed countries closest to what can be considered the global mainstream (e.g., the Netherlands, UK, and Belgium) and countries farthest from it (e.g., China, Iran, and Slovakia).
However, all of these works define culture in terms of country borders, which often makes sense, but is sometimes also problematic, for instance in countries with large minorities of inhabitants with different culture.

In our opinion, when building MRS, the analysis of cultural patterns of music consumption behavior, subsequent creation of respective cultural listener models, and their integration into recommender systems are vital steps to improve personalization and serendipity of recommendations.
Culture should be defined on various levels though, not only country borders. Other examples include having a joint historical background, speaking the same language, sharing the same beliefs or religion, and differences between urban vs.~rural cultures.
Another aspect that relates to culture is a temporal one since certain cultural trends, e.g., what defines the ``youth culture'', are highly dynamic in a temporal and geographical sense.
We believe that MRS which are aware of such cross-cultural differences and similarities in music perception and taste, and are able to recommend
music a listener in the same or another culture may like, would substantially benefit both users and providers of MRS.

\section{Conclusions}\label{sec:conclusions}
In this trends and survey paper, we identified several grand challenges the research field of music recommender systems (MRS) is facing. These are, among others, in the focus of current research in the area of MRS. 
We discussed (1) the \textit{cold start problem} of items and users, with its particularities in the music domain, (2) the challenge of \textit{automatic playlist continuation}, which is gaining importance due to the recently emerged user request of being recommended musical experiences rather than single tracks~\cite{schedl_etal:recsys:2017}, and (3) the challenge of holistically \textit{evaluating} music recommender systems, in particular, capturing aspects beyond accuracy.

In addition to the grand challenges, which are currently highly researched, we also presented a visionary outlook of what we believe to be the most interesting future research directions in MRS. In particular, we discussed (1) \textit{psychologically-inspired MRS}, which consider in the recommendation process factors such as listeners' emotion and personality, (2) \textit{situation-aware MRS}, which holistically model contextual and environmental aspects of the music consumption process, infer listener needs and intents, and eventually integrate these models at large scale in the recommendation process, and (3) \textit{culture-aware MRS}, which exploit the fact that music taste highly depends on the cultural background of the listener, where culture can be defined in manifold ways, including historical, political, linguistic, or religious similarities.

We hope that this article helped pinpointing major challenges, highlighting recent trends, and identifying interesting research questions in the area of music recommender systems.
Believing that research addressing the discussed challenges and trends will pave the way for the next generation of music recommender systems, we are looking forward to exciting, innovative approaches and systems that improve user satisfaction and experience, rather than just accuracy measures.





\section{acknowledgements}
We would like to thank all researchers in the fields of recommender systems, information retrieval, music research, and multimedia, with whom we had the pleasure to discuss and collaborate in recent years, and whom in turn influenced and helped shaping this article.
Special thanks go to Peter Knees and Fabien Gouyon for the fruitful discussions while preparing the ACM Recommender Systems 2017 tutorial on music recommender systems. 
In addition, we would like to thank the reviewers of our manuscript, who provided useful and constructive comments to improve the original draft and turn it into what it is now.
We would also like to thank Eelco Wiechert for providing additional pointers to relevant literature.
Furthermore, the many personal discussions with actual users of MRS unveiled important shortcomings of current approaches and in turn were considered in this article.



\end{document}